\documentclass[10pt,a4paper,onecolumn]{article}
\usepackage{marginnote}
\usepackage{graphicx}
\usepackage{xcolor}
\usepackage{authblk,etoolbox}
\usepackage{titlesec}
\usepackage{calc}
\usepackage{tikz}
\usepackage{hyperref}
\hypersetup{colorlinks,breaklinks=true,
            urlcolor=[rgb]{0.0, 0.5, 1.0},
            linkcolor=[rgb]{0.0, 0.5, 1.0}}
\usepackage{caption}
\usepackage{tcolorbox}
\usepackage{amssymb,amsmath}
\usepackage{ifxetex,ifluatex}
\usepackage{seqsplit}
\usepackage{xstring}

\usepackage{float}
\let\origfigure\figure
\let\endorigfigure\endfigure
\renewenvironment{figure}[1][2] {
    \expandafter\origfigure\expandafter[H]
} {
    \endorigfigure
}

\usepackage{fixltx2e} % provides \textsubscript
\usepackage[
  backend=biber,
]{biblatex}
\bibliography{paper.bib}

\let\textttOrig=\texttt
\def\texttt#1{\expandafter\textttOrig{\seqsplit{#1}}}
\renewcommand{\seqinsert}{\ifmmode
  \allowbreak
  \else\penalty6000\hspace{0pt plus 0.02em}\fi}

\makeatletter
\let\href@Orig=\href
\def\href@Urllike#1#2{\href@Orig{#1}{\begingroup
    \def\Url@String{#2}\Url@FormatString
    \endgroup}}
\def\href@Notdoi#1#2{\def\tempa{#1}\def\tempb{#2}%
  \ifx\tempa\tempb\relax\href@Urllike{#1}{#2}\else
  \href@Orig{#1}{#2}\fi}
\def\href#1#2{%
  \IfBeginWith{#1}{https://doi.org}%
  {\href@Urllike{#1}{#2}}{\href@Notdoi{#1}{#2}}}
\makeatother

\newlength{\cslhangindent}
\newlength{\csllabelwidth}
\newenvironment{CSLReferences}[3] % #1 hanging-ident, #2 entry spacing
 {% don't indent paragraphs
  \setlength{\parindent}{0pt}
  \ifodd #1 \everypar{\setlength{\hangindent}{\cslhangindent}}\ignorespaces\fi
  \ifnum #2 > 0
  \setlength{\parskip}{#2\baselineskip}
  \fi
 }%
 {}
\usepackage{calc}

\usepackage[top=3.5cm, bottom=3cm, right=1.5cm, left=1.0cm,
            headheight=2.2cm, reversemp, includemp, marginparwidth=4.5cm]{geometry}

\titleformat{\section}
  {\normalfont\sffamily\Large\bfseries}
  {}{0pt}{}
\titleformat{\subsection}
  {\normalfont\sffamily\large\bfseries}
  {}{0pt}{}
\titleformat{\subsubsection}
  {\normalfont\sffamily\bfseries}
  {}{0pt}{}
\titleformat*{\paragraph}
  {\sffamily\normalsize}

\usepackage{fancyhdr}
\makeatletter
\let\ps@plain\ps@fancy
\fancyheadoffset[L]{4.5cm}
\fancyfootoffset[L]{4.5cm}

\definecolor{linky}{rgb}{0.0, 0.5, 1.0}

\newtcolorbox{repobox}
   {colback=red, colframe=red!75!black,
     boxrule=0.5pt, arc=2pt, left=6pt, right=6pt, top=3pt, bottom=3pt}

\newcommand{\ExternalLink}{%
   \tikz[x=1.2ex, y=1.2ex, baseline=-0.05ex]{%
       \begin{scope}[x=1ex, y=1ex]
           \clip (-0.1,-0.1)
               --++ (-0, 1.2)
               --++ (0.6, 0)
               --++ (0, -0.6)
               --++ (0.6, 0)
               --++ (0, -1);
           \path[draw,
               line width = 0.5,
               rounded corners=0.5]
               (0,0) rectangle (1,1);
       \end{scope}
       \path[draw, line width = 0.5] (0.5, 0.5)
           -- (1, 1);
       \path[draw, line width = 0.5] (0.6, 1)
           -- (1, 1) -- (1, 0.6);
       }
   }

\patchcmd{\@maketitle}{center}{flushleft}{}{}
\patchcmd{\@maketitle}{center}{flushleft}{}{}
\patchcmd{\@maketitle}{\LARGE}{\LARGE\sffamily}{}{}
\def\maketitle{{%
  
  \AB@maketitle}}
\makeatletter
\renewcommand\AB@affilsepx{ \protect\Affilfont}
\renewcommand\AB@affilnote[1]{{\bfseries #1}\hspace{3pt}}
\renewcommand{\affil}[2][]%
   {\newaffiltrue\let\AB@blk@and\AB@pand
      \if\relax#1\relax\def\AB@note{\AB@thenote}\else\def\AB@note{#1}%
        \setcounter{Maxaffil}{0}\fi
        \begingroup
        \let\href=\href@Orig
        \let\texttt=\textttOrig
        \let\protect\@unexpandable@protect
        \def\thanks{\protect\thanks}\def\footnote{\protect\footnote}%
        \@temptokena=\expandafter{\AB@authors}%
        {\def\\{\protect\\\protect\Affilfont}\xdef\AB@temp{#2}}%
         \xdef\AB@authors{\the\@temptokena\AB@las\AB@au@str
         \protect\\[\affilsep]\protect\Affilfont\AB@temp}%
         \gdef\AB@las{}\gdef\AB@au@str{}%
        {\def\\{, \ignorespaces}\xdef\AB@temp{#2}}%
        \@temptokena=\expandafter{\AB@affillist}%
        \xdef\AB@affillist{\the\@temptokena \AB@affilsep
          \AB@affilnote{\AB@note}\protect\Affilfont\AB@temp}%
      \endgroup
       \let\AB@affilsep\AB@affilsepx
}
\makeatother

\renewcommand\Affilfont{\sffamily\small\mdseries}
\ifnum 0\ifxetex 1\fi\ifluatex 1\fi=0 % if pdftex
  \usepackage[T1]{fontenc}
  \usepackage[utf8]{inputenc}

\else % if luatex or xelatex
  \ifxetex
    \usepackage{mathspec}
    \usepackage{fontspec}

  \else
    \usepackage{fontspec}
  \fi
  \defaultfontfeatures{Ligatures=TeX,Scale=MatchLowercase}

\fi
\IfFileExists{upquote.sty}{\usepackage{upquote}}{}
\IfFileExists{microtype.sty}{%
\usepackage{microtype}
\UseMicrotypeSet[protrusion]{basicmath} % disable protrusion for tt fonts
}{}

\usepackage{hyperref}
\hypersetup{unicode=true,
            pdftitle={NEMESISPY: A Python package for simulating and retrieving exoplanetary spectra},
            pdfborder={0 0 0},
            breaklinks=true}
\let\addcontentslineOrig=\addcontentsline
\def\addcontentsline#1#2#3{\bgroup
  \let\texttt=\textttOrig\addcontentslineOrig{#1}{#2}{#3}\egroup}
\let\markbothOrig\markboth
\def\markboth#1#2{\bgroup
  \let\texttt=\textttOrig\markbothOrig{#1}{#2}\egroup}
\let\markrightOrig\markright
\def\markright#1{\bgroup
  \let\texttt=\textttOrig\markrightOrig{#1}\egroup}

\IfFileExists{parskip.sty}{%
\usepackage{parskip}
}{% else
\setlength{\parindent}{0pt}
\setlength{\parskip}{6pt plus 2pt minus 1pt}
}
\ifx\paragraph\undefined\else
\let\oldparagraph\paragraph
\renewcommand{\paragraph}[1]{\oldparagraph{#1}\mbox{}}
\fi
\ifx\subparagraph\undefined\else
\let\oldsubparagraph\subparagraph
\renewcommand{\subparagraph}[1]{\oldsubparagraph{#1}\mbox{}}
\fi

\title{NEMESISPY: A Python package for simulating and retrieving
exoplanetary spectra}

        \author[1]{Jingxuan Yang}
          \author[2]{Juan Alday}
          \author[1]{Patrick Irwin}
    
      \affil[1]{Department of Physics, University of Oxford, Parks Road,
Oxford OX1 3PU, UK}
      \affil[2]{School of Physical Sciences, The Open University, Walton
Hall, Milton Keynes MK7 6AA, UK}
  \date{\vspace{-7ex}}

\begin{document}
\maketitle

\marginpar{

  \begin{flushleft}
  %\hrule
  \sffamily\small

  {\bfseries DOI:} \href{https://doi.org/DOI unavailable}{\color{linky}{DOI unavailable}}

  \vspace{2mm}

  {\bfseries Software}
  \begin{itemize}
    \setlength\itemsep{0em}
    \item \href{N/A}{\color{linky}{Review}} \ExternalLink
    \item \href{NO_REPOSITORY}{\color{linky}{Repository}} \ExternalLink
    \item \href{DOI unavailable}{\color{linky}{Archive}} \ExternalLink
  \end{itemize}

  \vspace{2mm}

  \par\noindent\hrulefill\par

  \vspace{2mm}

  {\bfseries Editor:} \href{https://example.com}{Pending
Editor} \ExternalLink \\
  \vspace{1mm}
    {\bfseries Reviewers:}
  \begin{itemize}
  \setlength\itemsep{0em}
    \item \href{https://github.com/Pending Reviewers}{@Pending
Reviewers}
    \end{itemize}
    \vspace{2mm}

  {\bfseries Submitted:} N/A\\
  {\bfseries Published:} N/A

  \vspace{2mm}
  {\bfseries License}\\
  Authors of papers retain copyright and release the work under a Creative Commons Attribution 4.0 International License (\href{http://creativecommons.org/licenses/by/4.0/}{\color{linky}{CC BY 4.0}}).

  \end{flushleft}
}

\hypertarget{summary}{%
\section{Summary}\label{summary}}

Spectra of exoplanets allow us to probe their atmospheres' composition
and thermal structure and, when applicable, their surface conditions
(Burrows, 2014). Spectroscopic characterisation of a large population of
exoplanets may help us understand the origin and evolution of planetary
systems (Chachan et al., 2023; Nikku Madhusudhan et al., 2017; Mordasini
et al., 2016). The extraction of information from spectral data is known
as atmospheric retrievals (e.g., P. G. J. Irwin et al., 2008; Line et
al., 2013; N. Madhusudhan \& Seager, 2009), which can be divided into
two steps: forward modelling and model fitting. At a minimum, the
forward modelling step requires an atmospheric model for the observed
planet and a radiative transfer pipeline that can calculate model
spectra given some input atmospheric model. The model fitting step
typically requires a Bayesian parameter inference algorithm that can
constrain the free parameters of the forward model by fitting the
observed spectra. Atmospheric retrieval pipelines have long been applied
to the spectral analysis of the Earth and other solar system planets,
and the discovery of exoplanets further ignited the development of new
retrieval pipelines with varying focus and functionalities (MacDonald \&
Batalha, 2023).

NEMESISPY is a Python package developed to perform parametric
atmospheric modelling and radiative transfer calculation for the
retrievals of exoplanetary spectra. It is a recent development of the
well-established Fortran NEMESIS library (P. G. J. Irwin et al., 2008),
which has been applied to the atmospheric retrievals of both solar
system planets and exoplanets employing numerous different observing
geometries (J. K. Barstow et al., 2014, 2016; Joanna K. Barstow, 2020;
Patrick G. J. Irwin et al., 2020; James et al., 2023; Krissansen-Totton
et al., 2018; Lee et al., 2012; Teanby et al., 2012). NEMESISPY can be
easily interfaced with Bayesian inference algorithms to retrieve
atmospheric properties from spectroscopic observations. Recently,
NEMESISPY has been applied to the retrievals of Hubble and Spitzer data
of a hot Jupiter (Yang et al., 2023), as well as to JWST/Mid-Infrared
Instrument (JWST/MIRI) data of a hot Jupiter (Yang et al., 2024).

\hypertarget{statement-of-need}{%
\section{Statement of need}\label{statement-of-need}}

NEMESISPY has three distinguishing features as an exoplanetary retrieval
pipeline. Firstly, NEMESISPY inherits the fast correlated-k (Lacis \&
Oinas, 1991) radiative transfer routine from the Fortran NEMESIS library
(P. G. J. Irwin et al., 2008), which has been extensively validated
against other radiative transfer codes (Joanna K. Barstow et al., 2020).
Secondly, NEMESISPY employs a just-in-time compiler (Lam et al., 2015),
which compiles the most computationally expensive routines to machine
code at run time. Combined with extensive code refactoring, NEMESISPY is
significantly faster than the Fortran NEMESIS library. Such speed
improvement is crucial for analysing exoplanetary spectra using
sampling-based Bayesian parameter estimation (e.g., Feroz \& Hobson,
2008), which typically involves the computation of millions of model
spectra. Thirdly, NEMESISPY implements several parametric atmospheric
temperature models from Yang et al. (2023). These routines are
particularly useful for retrieving spectroscopic phase curves of hot
Jupiters, which are emission spectra observed at multiple orbital phases
and can enable detailed atmospheric characterisation.

NEMESISPY contains several general-purpose routines for atmospheric
modelling and spectral simulations. The modular nature of the package
means that subroutines can be easily called on their own. Currently,
NEMESISPY has an easy-to-use API for simulating emission spectra and
phase curves of hot Jupiters from arbitrary input atmospheric models,
and new features are being actively developed, such as multiple
scattering in radiative transfer calculation, an API for transmission
spectra, and the line-by-line radiative transfer method. NEMESISPY has
already enabled two scientific publications (Yang et al., 2024; Yang et
al., 2023) and is used in numerous ongoing exoplanetary data analysis
projects. The combination of well-tested core radiative transfer
routines, accelerated computational speed, and packaged modular design
is ideal for tackling the influx of JWST data of exoplanets.

\hypertarget{state-of-the-field}{%
\section{State of the field}\label{state-of-the-field}}

For a review of exoplanet atmospheric retrieval codes with comparable
functionalities to NEMESISPY, we refer the reader to the comprehensive
catalogue in MacDonald \& Batalha (2023).

\hypertarget{acknowledgements}{%
\section{Acknowledgements}\label{acknowledgements}}

The authors express gratitude to the developers of many open-source
Python packages used by NEMESISPY, in particular, numpy (Harris et al.,
2020), SciPy (Virtanen et al., 2020), Numba (Lam et al., 2015) and
Matplotlib (Hunter, 2007). The authors also express gratitude to the
many developers of the open-source Fortran NEMESIS library (P. G. J.
Irwin et al., 2008).

\hypertarget{references}{%
\section*{References}\label{references}}
\addcontentsline{toc}{section}{References}

\hypertarget{refs}{}
\begin{CSLReferences}{1}{0}
\leavevmode\hypertarget{ref-barstow_unveiling_2020}{}%
Barstow, Joanna K. (2020). Unveiling cloudy exoplanets: The influence of
cloud model choices on retrieval solutions. \emph{Monthly Notices of the
Royal Astronomical Society}, \emph{497}(4), 4183--4195.
\url{https://doi.org/10.1093/mnras/staa2219}

\leavevmode\hypertarget{ref-barstow_clouds_2014}{}%
Barstow, J. K., Aigrain, S., Irwin, P. G. J., Hackler, T., Fletcher, L.
N., Lee, J. M., \& Gibson, N. P. (2014). {CLOUDS ON THE HOT JUPITER
HD189733b}: {CONSTRAINTS FROM THE REFLECTION SPECTRUM}. \emph{The
Astrophysical Journal}, \emph{786}(2), 154.
\url{https://doi.org/10.1088/0004-637X/786/2/154}

\leavevmode\hypertarget{ref-barstow_consistent_2016}{}%
Barstow, J. K., Aigrain, S., Irwin, P. G. J., \& Sing, D. K. (2016). A
{CONSISTENT RETRIEVAL ANALYSIS OF} 10 {HOT JUPITERS OBSERVED IN
TRANSMISSION}. \emph{The Astrophysical Journal}, \emph{834}(1), 50.
\url{https://doi.org/10.3847/1538-4357/834/1/50}

\leavevmode\hypertarget{ref-barstow_comparison_2020}{}%
Barstow, Joanna K., Changeat, Q., Garland, R., Line, M. R., Rocchetto,
M., \& Waldmann, I. P. (2020). A comparison of exoplanet spectroscopic
retrieval tools. \emph{Monthly Notices of the Royal Astronomical
Society}, \emph{493}(4), 4884--4909.
\url{https://doi.org/10.1093/mnras/staa548}

\leavevmode\hypertarget{ref-burrows_spectra_2014}{}%
Burrows, A. S. (2014). Spectra as windows into exoplanet atmospheres.
\emph{Proceedings of the National Academy of Sciences}, \emph{111}(35),
12601--12609. \url{https://doi.org/10.1073/pnas.1304208111}

\leavevmode\hypertarget{ref-chachan_breaking_2023}{}%
Chachan, Y., Knutson, H. A., Lothringer, J., \& Blake, G. A. (2023).
Breaking {Degeneracies} in {Formation Histories} by {Measuring
Refractory Content} in {Gas Giants}. \emph{The Astrophysical Journal},
\emph{943}(2), 112. \url{https://doi.org/10.3847/1538-4357/aca614}

\leavevmode\hypertarget{ref-feroz_multimodal_2008}{}%
Feroz, F., \& Hobson, M. P. (2008). Multimodal nested sampling: An
efficient and robust alternative to {Markov Chain Monte Carlo} methods
for astronomical data analyses. \emph{Monthly Notices of the Royal
Astronomical Society}, \emph{384}(2), 449--463.
\url{https://doi.org/10.1111/j.1365-2966.2007.12353.x}

\leavevmode\hypertarget{ref-harris_array_2020}{}%
Harris, C. R., Millman, K. J., van der Walt, S. J., Gommers, R.,
Virtanen, P., Cournapeau, D., Wieser, E., Taylor, J., Berg, S., Smith,
N. J., Kern, R., Picus, M., Hoyer, S., van Kerkwijk, M. H., Brett, M.,
Haldane, A., del Río, J. F., Wiebe, M., Peterson, P., \ldots{} Oliphant,
T. E. (2020). Array programming with {NumPy}. \emph{Nature},
\emph{585}(7825), 357--362.
\url{https://doi.org/10.1038/s41586-020-2649-2}

\leavevmode\hypertarget{ref-hunter_matplotlib_2007}{}%
Hunter, J. D. (2007). Matplotlib: {A 2D Graphics Environment}.
\emph{Computing in Science \& Engineering}, \emph{9}(3), 90--95.
\url{https://doi.org/10.1109/MCSE.2007.55}

\leavevmode\hypertarget{ref-irwin_25d_2020}{}%
Irwin, Patrick G. J., Parmentier, V., Taylor, J., Barstow, J., Aigrain,
S., Lee, E., \& Garland, R. (2020). 2.{5D} retrieval of atmospheric
properties from exoplanet phase curves: Application to {WASP-43b}
observations. \emph{Monthly Notices of the Royal Astronomical Society},
\emph{493}(1), 106--125. \url{https://doi.org/10.1093/mnras/staa238}

\leavevmode\hypertarget{ref-irwin_nemesis_2008}{}%
Irwin, P. G. J., Teanby, N. A., de Kok, R., Fletcher, L. N., Howett, C.
J. A., Tsang, C. C. C., Wilson, C. F., Calcutt, S. B., Nixon, C. A., \&
Parrish, P. D. (2008). The {NEMESIS} planetary atmosphere radiative
transfer and retrieval tool. \emph{Journal of Quantitative Spectroscopy
and Radiative Transfer}, \emph{109}(6), 1136--1150.
\url{https://doi.org/10.1016/j.jqsrt.2007.11.006}

\leavevmode\hypertarget{ref-james_temporal_2023}{}%
James, A., Irwin, P. G. J., Dobinson, J., Wong, M. H., Tsubota, T. K.,
Simon, A. A., Fletcher, L. N., Roman, M. T., Teanby, N. A., Toledo, D.,
\& Orton, G. S. (2023). The {Temporal Brightening} of {Uranus}'
{Northern Polar Hood From HST}/{WFC3} and {HST}/{STIS Observations}.
\emph{Journal of Geophysical Research: Planets}, \emph{128}(10),
e2023JE007904. \url{https://doi.org/10.1029/2023JE007904}

\leavevmode\hypertarget{ref-krissansen-totton_detectability_2018}{}%
Krissansen-Totton, J., Garland, R., Irwin, P., \& Catling, D. C. (2018).
Detectability of {Biosignatures} in {Anoxic Atmospheres} with the james
webb space telescope: {A TRAPPIST-1e Case Study}. \emph{The Astronomical
Journal}, \emph{156}(3), 114.
\url{https://doi.org/10.3847/1538-3881/aad564}

\leavevmode\hypertarget{ref-lacis_description_1991}{}%
Lacis, A. A., \& Oinas, V. (1991). A description of the correlated
{\emph{k}} distribution method for modeling nongray gaseous absorption,
thermal emission, and multiple scattering in vertically inhomogeneous
atmospheres. \emph{Journal of Geophysical Research}, \emph{96}(D5),
9027. \url{https://doi.org/10.1029/90JD01945}

\leavevmode\hypertarget{ref-lam_numba_2015}{}%
Lam, S. K., Pitrou, A., \& Seibert, S. (2015). Numba: A {LLVM-based
Python JIT} compiler. \emph{Proceedings of the {Second Workshop} on the
{LLVM Compiler Infrastructure} in {HPC}}, 1--6.
\url{https://doi.org/10.1145/2833157.2833162}

\leavevmode\hypertarget{ref-lee_optimal_2012}{}%
Lee, J.-M., Fletcher, L. N., \& Irwin, P. G. J. (2012). Optimal
estimation retrievals of the atmospheric structure and composition of
{HD} 189733b from secondary eclipse spectroscopy: {Exoplanet} retrieval
from transit spectroscopy. \emph{Monthly Notices of the Royal
Astronomical Society}, \emph{420}(1), 170--182.
\url{https://doi.org/10.1111/j.1365-2966.2011.20013.x}

\leavevmode\hypertarget{ref-line_systematic_2013}{}%
Line, M. R., Wolf, A. S., Zhang, X., Knutson, H., Kammer, J. A.,
Ellison, E., Deroo, P., Crisp, D., \& Yung, Y. L. (2013). A {SYSTEMATIC
RETRIEVAL ANALYSIS OF SECONDARY ECLIPSE SPECTRA}. {I}. {A COMPARISON OF
ATMOSPHERIC RETRIEVAL TECHNIQUES}. \emph{The Astrophysical Journal},
\emph{775}(2), 137. \url{https://doi.org/10.1088/0004-637X/775/2/137}

\leavevmode\hypertarget{ref-macdonald_catalog_2023}{}%
MacDonald, R. J., \& Batalha, N. E. (2023). A {Catalog} of {Exoplanet
Atmospheric Retrieval Codes}. \emph{Research Notes of the AAS},
\emph{7}(3), 54. \url{https://doi.org/10.3847/2515-5172/acc46a}

\leavevmode\hypertarget{ref-madhusudhan_atmospheric_2017}{}%
Madhusudhan, Nikku, Bitsch, B., Johansen, A., \& Eriksson, L. (2017).
Atmospheric signatures of giant exoplanet formation by pebble accretion.
\emph{Monthly Notices of the Royal Astronomical Society}, \emph{469}(4),
4102--4115. \url{https://doi.org/10.1093/mnras/stx1139}

\leavevmode\hypertarget{ref-madhusudhan_temperature_2009}{}%
Madhusudhan, N., \& Seager, S. (2009). A {TEMPERATURE AND ABUNDANCE
RETRIEVAL METHOD FOR EXOPLANET ATMOSPHERES}. \emph{The Astrophysical
Journal}, \emph{707}(1), 24--39.
\url{https://doi.org/10.1088/0004-637X/707/1/24}

\leavevmode\hypertarget{ref-mordasini_imprint_2016}{}%
Mordasini, C., van Boekel, R., Mollière, P., Henning, Th., \& Benneke,
B. (2016). {THE IMPRINT OF EXOPLANET FORMATION HISTORY ON OBSERVABLE
PRESENT-DAY SPECTRA OF HOT JUPITERS}. \emph{The Astrophysical Journal},
\emph{832}(1), 41. \url{https://doi.org/10.3847/0004-637X/832/1/41}

\leavevmode\hypertarget{ref-teanby_active_2012}{}%
Teanby, N. A., Irwin, P. G. J., Nixon, C. A., de Kok, R., Vinatier, S.,
Coustenis, A., Sefton-Nash, E., Calcutt, S. B., \& Flasar, F. M. (2012).
Active upper-atmosphere chemistry and dynamics from polar circulation
reversal on {Titan}. \emph{Nature}, \emph{491}(7426), 732--735.
\url{https://doi.org/10.1038/nature11611}

\leavevmode\hypertarget{ref-virtanen_scipy_2020}{}%
Virtanen, P., Gommers, R., Oliphant, T. E., Haberland, M., Reddy, T.,
Cournapeau, D., Burovski, E., Peterson, P., Weckesser, W., Bright, J.,
van der Walt, S. J., Brett, M., Wilson, J., Millman, K. J., Mayorov, N.,
Nelson, A. R. J., Jones, E., Kern, R., Larson, E., \ldots{} van
Mulbregt, P. (2020). {SciPy} 1.0: Fundamental algorithms for scientific
computing in {Python}. \emph{Nature Methods}, \emph{17}(3), 261--272.
\url{https://doi.org/10.1038/s41592-019-0686-2}

\leavevmode\hypertarget{ref-yang_simultaneous_2024}{}%
Yang, J., Hammond, M., Piette, A. A. A., Blecic, J., Bell, T. J., Irwin,
P. G. J., Parmentier, V., Tsai, S.-M., Barstow, J. K., Crouzet, N.,
Kreidberg, L., Mendonça, J. M., Taylor, J., Baeyens, R., Ohno, K.,
Teinturier, L., \& Nixon, M. C. (2024). Simultaneous retrieval of
orbital phase resolved {JWST}/{MIRI} emission spectra of the hot
{Jupiter WASP-43b}: {Evidence} of water, ammonia and carbon monoxide.
\emph{Monthly Notices of the Royal Astronomical Society}, stae1427.
\url{https://doi.org/10.1093/mnras/stae1427}

\leavevmode\hypertarget{ref-yang_testing_2023}{}%
Yang, J., Irwin, P. G. J., \& Barstow, J. K. (2023). Testing {2D}
temperature models in {Bayesian} retrievals of atmospheric properties
from hot {Jupiter} phase curves. \emph{Monthly Notices of the Royal
Astronomical Society}, \emph{525}(4), 5146--5167.
\url{https://doi.org/10.1093/mnras/stad2555}

\end{CSLReferences}

\end{document}